\input psfig

\catcode`@=11
\def\leftrightarrowfill{$\m@th \mathord\leftarrow\mkern-6mu
\cleaders\hbox{$\mkern-2mu\mathord-\mkern-2mu$}\hfill
\mkern-6mu\mathord\rightarrow$}
\catcode`@=12
\def\overlrarrow#1{\vbox{\ialign{##\crcr\leftrightarrowfill\crcr\noalign
{\kern-1pt\nointerlineskip}$\hfil\displaystyle{#1}\hfil$\crcr}}}

\documentstyle[11pt]{article}

\begin{document}
\vspace*{-1.25in}
\vspace*{.85in}
\begin{center}
{\Large{\bf Pion Wavefunctions and Truncation Sensitivity of QCD Sum Rules}} \\
\vspace*{.45in}
{\large{A.~Duncan$^1$, S.~Pernice$^2$, and
E.~Schnapka$^3$}} \\ 
\vspace*{.15in}
$^1$Dept. of Physics, Columbia University, New York, NY 10027\\
(permanent address: Dept. of Physics and Astronomy, Univ. of Pittsburgh, Pittsburgh, PA 15620)\\
$^2$Dept. of Physics and Astronomy, Univ. of Rochester, Rochester, NY 14627\\
$^3$Institut f\"ur theoretische Physik, Technische Universit\"at, M\"unchen,\\
 James-Franck-Strasse, D-85748 Garching, Germany\\
\end{center}
e-mail: tony@dectony.phyast.pitt.edu, sergio@charm.pas.rochester.edu,
schnapka@hertz.t30.physik.tu-muenchen.de
\vspace*{.3in}
\begin{abstract}
The systematic errors inherent in the QCD sum rule approach to 
meson wavefunctions are examined in the context of QCD in
1+1 spacetime dimensions in the large N limit where the
theory is exactly solvable. It is shown that the truncation of
high momentum modes induced in a lattice discretization 
automatically produces a Chernyak-Zhitnitsky \cite{CZ} type meson
wavefunction. Such a truncation alters the balance of leading and
higher twist terms in correlators.  We find that the reliable extraction of  (a few) higher moments
is possible provided a reasonably accurate uniform approximation to
the Euclidean correlator over a suitable $Q^2$ range is available, but
that the extracted values are particularly sensitive to the balance of lower and
higher twist contributions. Underestimates of lower twist contributions or
overestimates of the highest twist term may lead to  too high values for the second
and fourth moments of the pion wavefunction, suggesting a doubly peaked
structure of the Chernyak-Zhitnitsky type.
\end{abstract}


\newpage
\section{Introduction}
  
  Although lattice gauge theory remains the principal technique for
extracting reliable nonperturbative information in 4 dimensional
quantum chromodynamics (QCD), the method of QCD sum rules \cite{qcdsr},
 in which physical results are extracted by combining perturbation
theory with nonperturbative information encapsulated in a small number
of hadronic matrix elements (``condensates"), has also enjoyed 
considerable popularity. One of the more startling (and certainly
unanticipated) results obtained by the sum rules technique is the
prediction of Chernyhak and Zhitnitsky \cite{CZ} that the pion wavefunction
in light-cone gauge has a doubly peaked, non-convex structure in which
one of the quarks is most likely to be carrying the preponderance of the
meson momentum. Some early lattice calculations \cite{Kron} found
even larger values for the second moment, restoring the maximum
probability to equally shared momenta, but producing {\em negative}
values for the ground state wavefunction. Although a double peak wavefunction
gives 
a better fit to the experimental data on the pion form factor at 
accessible energies, such an Ansatz has come under increasing scrutiny \cite{AMN},
\cite{LSR},
as it is unclear whether higher twist contributions (i.e. higher Fock
states than simply a valence quark-antiquark) are really negligible in
the accessible $Q^2$ range, in particular in the end-point regions which
are necessarily emphasized by a wavefunction of Chernyak-Zhitnitsky type.

  In this paper we report the results of a careful study of the systematic
errors intrinsic to the sum rules approach in an essentially solvable model
with close resemblance to 4 dimensional QCD, namely, the large N (= number
of colors) limit of two-dimensional quantum chromodynamics. 
  Quantized gauge theories become particularly simple in two
space-time dimensions, a feature first exploited by Schwinger\cite{Schw}
 in
his seminal paper on massless quantum electrodynamics. The
analytic tractability of the nonabelian theory with gauge group
SU(N) in the limit of large N was utilized by 't Hooft\cite{tHooft} in his study of
the meson spectrum in 2-dimensional QCD. It turns out that this model
allows us to study in great detail the accuracy of the sum rule method
and its sensitivity to various types of systematic errors in the input
data in a situation where the correlators being studied and the spectrum
and physical matrix elements which are essential ingredients of the
procedure are under complete analytic or numerical control. 

   One clue to the type of systematic error which can wreak havoc in
a sum rules approach is provided by the observation that the CZ
wavefunction appears automatically in a rather simple truncation of
the theory: namely, in the boost of the lattice discretized Coulomb
gauge wavefunction to light-cone gauge. In Section 2 we explain how this
comes about by showing how to connect a relativistic quark hamiltonian
appropriate for an equal-time formalism with the 't Hooft Hamiltonian
appropriate for light-cone gauge. In Section 3 we review and extend some
well-known properties of 2-dimensional QCD at large N. Section 4 contains
some 1 and 2-loop perturbative calculations which allow us to compute
the required higher twist terms in the asymptotic expansion of the
correlators studied in the sum rules approach to wavefunctions. In
Section 5 we show how the method can be applied successfully, provided
a reasonably accurate uniform fit to the Euclidean correlator is available
over a suitable $Q^2$ range. Precise control of logarithmic terms in 
the higher twist parts is found not to be crucial (although certainly more
accurate results are obtained if we include these terms correctly). 
In Section 6, the method is shown to fail seriously if the balance
of lower and higher twist terms is altered (as it is in the lattice 
discretized situation discussed in Section 2). This may happen simply because
the higher dimension condensates depend on a high power of the QCD scale,
which is not known with precision, or if unexpected nonperturbative
terms (such as the infamous first infrared renormalon \cite{BYZ}) 
should happen to be present.



\section{Light-Cone Wavefunctions in Lattice discretized QCD}

The bound state equation in 2 dimensional QCD in Coulomb gauge 
and in the limit $N_c \rightarrow \infty$ 
has been thoroughly investigated by Bars and Green \cite{bars}.
Up to a unitary transformation the full Bethe--Salpeter 
wavefunction can be separated into two contributions 
\begin{equation}
\phi = \phi_+ \frac{(1+\gamma_0)}{2}\gamma_5 
     + \phi_- \frac{(1-\gamma_0)}{2}\gamma_5 
\label{eq:phi}
\end{equation}
corresponding to a $q \bar{q}$ pair moving forward and backward in time. 
$\phi_+$ and $\phi_-$ are defined by a coupled set of integral equations, 
\begin{eqnarray}
\lefteqn{ [E(p)+E(r-p)-r^0] \phi_+ (r,p) =} \nonumber \\
 & & \frac{g^2 N}{2\pi} P \int 
     \frac{dk}{(p-k)^2} [C(p,k,r)\phi_+ (r,k) + S(p,k,r)\phi_- (r,k)]  
     \label{eq:bse1} \\
\lefteqn{ [E(p)+E(r-p)+r^0] \phi_- (r,p) =} \nonumber \\
 & & \frac{g^2 N}{2\pi} P \int 
     \frac{dk}{(p-k)^2} [C(p,k,r)\phi_- (r,k) + S(p,k,r)\phi_+ (r,k)] 
     \label{eq:bse2}\;\;,
\end{eqnarray}
where 
\begin{eqnarray}
C(p,k,r) & = & \cos \frac{1}{2} [\Theta(p)-\Theta(k)]
               \cos \frac{1}{2} [\Theta(r-p)-\Theta(r-k)] \;\;, \\
S(p,k,r) & = & \sin \frac{1}{2} [\Theta(p)-\Theta(k)]
               \sin \frac{1}{2} [\Theta(r-k)-\Theta(r-p)] \;\;.
\end{eqnarray}
The functions $E(p)$ and $\Theta(p)$ are parametrizations of the quark 
self energy  
\begin{equation}
\label{eq:sigma}
\Sigma(p) = [E(p)\cos\Theta(p)-m] + \gamma^1 [E(p)\sin\Theta(p)-p]
\end{equation}
and are determined by the self-consistent quark self energy equation.
One can easily solve for $E(p)$ and $\Theta(p)$ in the nonrelativistic 
(small coupling) limit, where eqs.(\ref{eq:bse1}) and (\ref{eq:bse2}) 
decouple and yield the ordinary non-relativistic Schr\"odinger equation. 
Likewise one can solve for $E(p)$ and $\Theta(p)$ in the large momentum 
limit, finding 
\begin{eqnarray}
\Theta(p) & = & \frac{p}{|p|}
        \left[ \frac{\pi}{2} - \frac{m}{p} + O (\frac{1}{p^2}) \right] \\
E(p)      & = & |p| - \frac{g^2 N/2\pi}{|p|} + \frac{m^2}{2|p|} 
   + O (\frac{1}{p^2})  \;\;. 
\end{eqnarray}
Bars and Green pointed out   that in the boosted limit $r \rightarrow \infty$ 
and $ xp = r$ the coupled equations (\ref{eq:bse1},\ref{eq:bse2}) decouple as well,  
yielding a vanishing $\phi_-$ and one remaining  integral equation for 
$\phi_+$ similar to 't~Hooft's integral equation \cite{tHooft} obtained 
in light-cone gauge. 
In \cite{thesis} it was shown how  
a Green's function in Coulomb gauge can be analytically transformed 
to light-cone gauge. This involves boosting the momenta of the wavefunctions
and performing appropriate transformations in Dirac space to obtain the 
correct $\gamma$ matrix structure. The transformation formula reads 
\begin{equation}
S_\Lambda \Psi_{Coulomb}(\Lambda^{-1}r,\Lambda^{-1}p) S_\Lambda^{-1} 
 = \Psi_{(\omega)}(r,p) \stackrel{\omega \rightarrow \infty}{\longrightarrow}
   \Psi_{Lightcone}(r,p)\;,
\label{eq:trans}
\end{equation}
with the Lorentz transformation
\begin{equation}
\Lambda = \left( 
 \begin{array}{rr}
  \cosh(\omega) & -\sinh(\omega) \\
 -\sinh(\omega) &  \cosh(\omega)
 \end{array} \right)
\end{equation}
and the corresponding spinor matrix $S_\Lambda$. Note that 
\begin{equation}
\Lambda^{-1}p \stackrel{\omega \rightarrow \infty}{\longrightarrow}
  \frac{\exp(\omega)}{\sqrt{2}}\left(
 \begin{array}{ll}
  p_- \\
  p_- 
 \end{array}\right)
\end{equation}
and the transformation (\ref{eq:trans}) produces the dependence of the light-cone 
Green's function on light-cone variables as expected. All the factors of 
$\exp(\omega)$ which diverge in the limit $\omega \rightarrow \infty$ 
cancel in the full transformation of the Bethe--Salpeter
equation 
(\ref{eq:phi}),(\ref{eq:bse1}),(\ref{eq:bse2}) to the 't~Hooft light-cone equation 
according to  (\ref{eq:trans}). 

For our purposes we may  introduce a single relativistic Schr\"odinger 
equation (relativistic quark model) which intuitively describes the 
binding of two quarks in a linear rising potential:
\begin{equation}
\label{eq:relSchr}
 {\cal H} \phi(x) \equiv \left(
 \sqrt{\hat{p}^{2}+m_a^{2}}+\sqrt{(\hat{r}-\hat{p})^{2}+m_b^{2}}-m_a-m_b
+\frac{\pi}{2}|x| \right) \phi(x) = \epsilon \phi(x)\;\;.
\end{equation}
This equation interpolates properly between the nonrelativistic limit (for $r=0$) and
the boosted high momentum limit of the exact Bethe--Salpeter 
equation (\ref{eq:bse1}),(\ref{eq:bse2}). In either case $\phi_-$ vanishes 
while equation (\ref{eq:bse1}) for $\phi_+$ plays the central  role (with
$C=1,\;S=0$).

In (\ref{eq:relSchr}) the prefactor $g^2N/\pi$ from the potential is removed 
and all masses and momenta are measured in units of the 
coupling constant $g\sqrt(N/\pi)$, likewise distances in units of 
 $(g\sqrt(N/\pi))^{-1}$. 
%

In the following the relativistic quark model (\ref{eq:relSchr}) is treated   
by discretization on a lattice. Plotting the resulting 
wavefunction in terms of the variable $x=p/r$ yields the approximate 
(and in the weak coupling limit, the exact)  
light-cone result. The boost is accounted for 
by using a large value of total momentum $r$, meaning $r \gg m_a, m_b$. 

We perform the discretization in coordinate space before going back to 
momentum space by Fast Fourier Transformation. Introducing a lattice 
spacing $a$, the physical box size $L=Na$ and $i_p$ as the integer 
lattice momentum corresponding to $p$, the latticized equation reads
\begin{eqnarray}
\lefteqn{ \Bigg( 
 \sqrt{\left(\frac{2}{a}\sin \frac{\pi i_p}{N}\right)^2 +m_a^{2} }
+\sqrt{\left(\frac{2}{a}\sin \frac{\pi (i_r-i_p)}{N}\right)^2 + m_b^{2} }
   } \\
 & &  -m_a-m_b \Bigg) \tilde{\phi}_{i_p} 
 + \frac{1}{N} \sum_{i_q,n=0}^{\infty} \frac{\pi}{2}a \min(n,N-n) 
  e^{2\pi i(i_p-i_q)n/N} \tilde{\phi}_{i_q} = \epsilon \tilde{\phi}_{i_p}\;\;.
\end{eqnarray}
As a result of the discretization, the differential operator 
$i \partial/\partial x$ corresponds to the momentum space operator 
$(2/a)\sin(\pi i_p/N)$ instead of to the naive lattice momentum 
$2\pi i_p/L$. This difference becomes important for physical momenta
(in our case the total momentum $r$) approaching the UV cut--off $2/a$. 

We choose lattice sizes of $N=512, N=1024$ and $N=2048$ to investigate the
ground state wavefunction for equal quark masses  $m=1$  
 shown in Fig(1).
The value of $r$ as well as the {\em physical}
box size $L$ is kept fixed for each value of $m$, such that with decreasing 
number of points $N$ the physical lattice spacing $a=L/N$ increases and the 
resulting UV momentum cut-off decreases. The result is rather startling:
as the ultraviolet cutoff (inverse lattice spacing) is reduced the
characteristic convex, singly-peaked form of the ground-state meson
wavefunction found by 't Hooft is replaced by a double peaked structure
with a {\em minimum} at $x=1/2$ - exactly the shape suggested by
the QCD sum rule method of Chernyak and Zhitnitsky \cite{CZ}, or
 indeed, by some early lattice results \cite{Kron}! The
cutoff of quark high momentum modes alters completely the balance 
of leading and subleading terms in the large $Q^2$ behavior of 
correlators of quark-antiquark currents, of course, and we shall see
below that it is precisely such an alteration that can lead to large
systematic errors in the extraction of moments of meson wavefunctions
in the sum rule approach.
\begin{figure}[htp]
\hbox to \hsize{\hss\psfig{figure=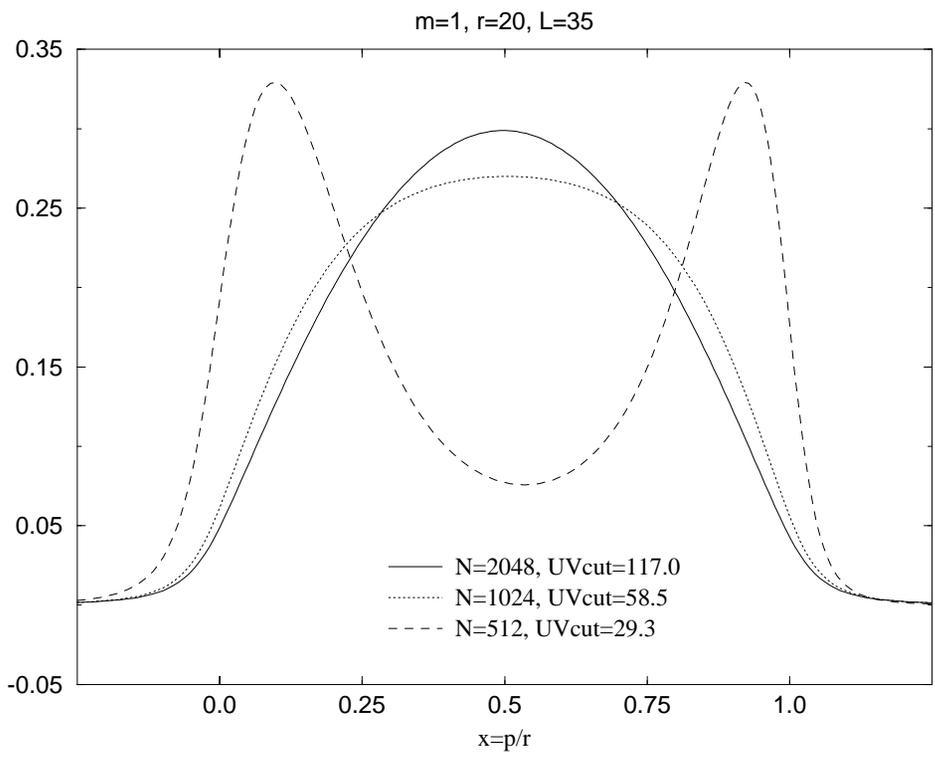,width=0.8\hsize}\hss}
\caption{Light Cone wavefunctions with lattice discretization ($m=1$)}
\label{fig:fig1)}
\end{figure}

\section{Review of some results from 2-dimensional QCD}

 In the  limit where the number of colors $N$ is taken large,
with $g^2 N/\pi$ held fixed, QCD in 1 space-1 time dimension
exhibits a discrete
spectrum of stable mesons of mass $\mu_k$ ($k$=1,2,..), and duality is
exact in the sense that 2 point correlators are meromorphic functions
of $q^2$ expressible as sums over resonance poles. As pointed out
by Callan et al.\cite{CCG}, the large $q^2$ behavior arising from the asymptotic
freedom of the theory is intimately related to the asymptotic behavior
for large $k$ (meson excitation level) of the meson masses $\mu_k$
and decay constants $f_k$.

   A convenient test case for examining the sensitivity of the sum rule approach
to truncations of the short distance expansion  is provided by the two point
 correlators of  the tower of light-cone operators
\begin{equation}
\label{eq:sn}
   S_n(x) \equiv \frac{2}{\sqrt{N}}\bar{\psi}(x) \gamma_5 (\overlrarrow{D}_-)^n\psi(x)
\end{equation}
which reduce in light-cone gauge ($A_-$=0) to
\begin{equation}
\label{eq:snax}
    S_n(x)=\frac{2}{\sqrt{N}}\bar{\psi}(x) \gamma_5 (\overlrarrow{\partial}_-)^n\psi(x)
\end{equation}
The standard sum rule approach to the pion wavefunction due to Chernyak and
Zhitnitsky \cite{CZ} begins with a short distance expansion for a two point correlator 
of  currents (or densities) which couple to the meson in question. It turns out
that the tower of operators based on a pseudoscalar density has the same
leading behavior for large $q^2$ in 2 dimensions as the two-point correlator
of axial currents in 4 dimensional QCD. Hence, we study
\begin{eqnarray}
 M_{n}(q^2)&\equiv&\frac{-i}{q_{-}^n}\int d^{2}x e^{iqx}<0| T\{S_n(x)S_0(0)|0> -(q^2\rightarrow -\mu^2) \\
     &\approx& \frac{2}{\pi}\ln(\frac{-q^2}{\mu^2})+O(1/q^2),\;\;Q^2\equiv -q^2\rightarrow\infty
\end{eqnarray}

This correlator may also be written (in the large N limit) as a sum over
resonance poles
\begin{eqnarray}
\label{eq:ressum}
  M_{n}(q^2)&=&\sum_{k\;\rm odd}\frac{<0|S_n(0)|k><k|S_0(0)|0>}{q^2-\mu_{k}^2} -(q^2\rightarrow -\mu^2) \nonumber \\
  &=&\sum_{k\;\rm odd}\frac{f_{nk}f_{0k}}{q^2-\mu_{k}^2}-(q^2\rightarrow -\mu^2)
\end{eqnarray}
where the squared meson masses are eigenvalues of the 't Hooft Hamiltonian
\begin{equation}
\label{eq:Ham}
 H\phi_{k}(x)=\frac{\gamma}{x(1-x)}\phi_{k}(x)+P\int_{0}^{1}\frac{\phi_{k}(x)-\phi_{k}(y)}{(x-y)^2}dy =\mu_k^2\phi_{k}(x)
\end{equation}
Here the parameter $\gamma$ is the bare quark mass  (which we take equal to the antiquark mass
throughout) squared in units where $\frac{g^2 N}{\pi}$ is set to unity.  The resonance residues are
quadratic in the moments
\begin{equation}
 f_{nk}\equiv \int_{0}^{1}\frac{(1-2x)^{n}}{x(1-x)}\phi_{k}(x)dx
\end{equation}
with the normalization
\begin{equation}
  \int_{0}^{1} \phi_{k}^2(x)dx = 1
\end{equation}
As shown in \cite{CCG}, for any given moment
\begin{equation}
  f_{nk}\rightarrow \frac{2\pi}{\sqrt{\gamma}}(1+\frac{A_n}{k}+O(\frac{1}{k^2}\ln(k))),\;\;k\rightarrow\infty
\end{equation}
which, together with the asymptotic behavior derived originally by 't Hooft
\cite{tHooft}
\begin{equation}
  \mu_{k}^2 \simeq \pi^2 k+2(\gamma-1)\ln(k)+O(1/k)
\end{equation}
implies the required free field behavior  of  $M_n(q^2)$ at large $Q^2$. 

  The meson wavefunctions $\phi_{k}(x)$ are not known analytically, but may be obtained to
high accuracy numerically by expanding in a finite basis:
\begin{equation}
\label{eq:basis}
  \phi_{k}(x)=c_{k0}x^{\beta}(1-x)^{\beta}+\sqrt{2}\sum_{j=1}^{D}c_{kj}\sin(\pi (2j-1)x)
\end{equation}
The parameter $\beta$ is related to the quark mass by $\pi\beta/\tan(\pi\beta)=1-\gamma$. The first
basis function appearing on the right-hand-side of (\ref{eq:basis}) is necessary to properly treat the nonanalytic 
behavior at the endpoints x=0,1. In the above basis the 't Hooft Hamiltonian becomes a
($D$+1)x($D$+1) matrix which may be numerically diagonalized by standard
techniques.  Adequate accuracy  (typically to 5 significant places at least) in
all the quantities computed from these eigenfunctions was obtained by taking $D$=120. The
groundstate pseudoscalar meson (``pion") wavefuntions for a range of quarkmasses
($m_q^2=$0.135,0.534,1.0,2.60) are shown in Fig.2. These wavefunctions are in all cases
convex, with a single extremum at $x=\frac{1}{2}$.


\begin{figure}[htp]
\hbox to \hsize{\hss\psfig{figure=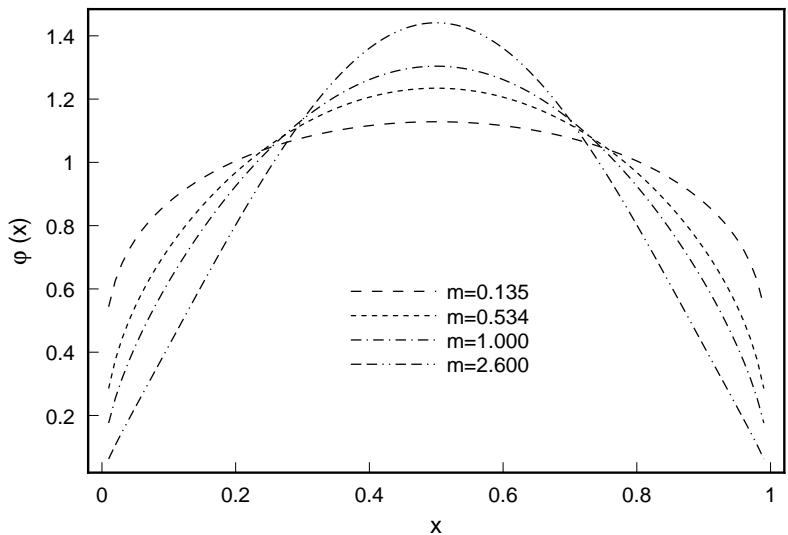,width=0.8\hsize}\hss}
\caption{Ground state (pion) wavefunctions in 2D QCD}
\label{fig:phi(x)}
\end{figure}

   Once the $\phi_{k}(x)$ have been computed, accurate values for the moments $f_{nk}$
are readily obtained by numerical integration. The nonanalytic singularities at $x=0,1$
are best avoided by using the following exact identities, which follow from the integral
equation (\ref{eq:Ham})
\begin{eqnarray}
  f_{0k}&=&\frac{\mu_k^2}{\gamma}\int_{0}^{1}\phi_{k}(x)dx \nonumber \\
     &\simeq&\frac{2\pi}{\sqrt{\gamma}}(1+\frac{A_0}{k}+O(\frac{1}{k^2}\ln(k))),\;\;k\rightarrow\infty
\end{eqnarray}
where the coefficient $A_0$,  which is related to the higher twist behavior of $M_0(q^2)$,
 will be determined analytically below. The second moment  is determined in terms of
the zeroth as follows (we use the property $\phi_{k}(x)=\phi_{k}(1-x)$ for
the pseudoscalar mesons)
\begin{eqnarray}
\label{eq:identity2}
 f_{2k}&=&\int_{0}^{1}\frac{1}{x(1-x)}(1-2x)^2\phi_{k}(x)dx \nonumber \\
   &=& 2\int_{0}^{1}(\frac{1}{x}-4+4x)\phi_{k}(x)dx \nonumber \\
  &=&(1-\frac{4\gamma}{\mu_k^2})f_{0k}  \nonumber \\
  &\simeq&\frac{2\pi}{\sqrt(\gamma)}(1+(A_0-\frac{4\gamma}{\pi^2})\frac{1}{k}+O(\frac{1}{k^2}\ln(k)))
\end{eqnarray}
while $f_{4k}$ may be related  to $f_{0k}$ and a nonsingular integral which may be readily
computed to high accuracy once the wavefunctions are known:
\begin{equation}
\label{eq:identity4}
   f_{4k}=f_{0k}-16\sqrt{\frac{\gamma}{\pi}}\int_{0}^{1}x^2\phi_{k}(x)dx
\end{equation}
In the sum rules approach, the first three moments ($f_{0k},f_{2k},$ and $f_{4k}$) are estimated
and then used to draw conclusions about the shape of the pion wavefunction. We shall therefore
restrict our attention to these quantities below. 

   The subdominant terms in the asymptotic expansion for large $k$ of $f_{nk}$ are 
related to {\em logarithmic} higher twist terms in $M_n(q^2)$. For example, it follows
from the resonance sum representation of $M_0(q^2)$ that, as $Q^2 \equiv -q^2\rightarrow  +\infty$
\begin{equation}
 M_{0}(q^2)\simeq \frac{2}{\pi}\ln{\frac{Q^2}{\mu^2}}+\frac{4}{\pi}\frac{\gamma-1-\pi^2 A_0}{Q^2}\ln{\frac{Q^2}{\mu^2}}+O(1/Q^2)
\end{equation}
 The pure $1/Q^2$ however (without a logarithm) is {\em not} determined by the large
k asymptotics. Indeed, dropping a finite number of initial terms in (\ref{eq:ressum})  clearly alters the
coefficient of $1/Q^2$, while leaving the large k behavior unchanged.  In the following section
we will compute the next to leading twist contributions to $M_n(q^2)$ directly from
2 loop perturbation theory. This is possible as a consequence of the superrenormalizability
of the theory. 

  The sum rules technique involves, as mentioned above,  an approximate determination
of the first three moments of the groundstate pseudoscalar . In  2-dimensional QCD, the
 moments $f_{nk}$ vanish by parity in the pseudoscalar sector generated 
by the basis (\ref{eq:basis})  unless $n$ is even.
Thus we will be computing the quantities $f_{01},f_{21}$ and $f_{41}$. In the large N
limit they are obtained essentially exactly by the numerical procedure outlined above.
\section{Large $Q^2$ behavior of Correlators}

  In the large N limit of 2 dimensional QCD, a superrenormalizable theory, the leading behavior 
of the correlators $M_n(q^2)$ in the deep Euclidean regime $-q^2\equiv Q^2\rightarrow\infty$
is given by the single one-loop graph (Fig(3a)) in which the quark-antiquark pair propagates
freely . Keeping quark mass dependent terms, the Feynman integral is readily performed
for general $n$ and one finds at large $Q^2$
\begin{eqnarray}
\label{eq:1loop}
 M_n^{\rm 1-loop}(Q^2)&\simeq& \frac{2}{\pi}\ln{\frac{Q^2}{m^2}}+\frac{4}{\pi}(n-1)\frac{m^2}{Q^2}\ln{\frac{Q^2}{m^2}} \nonumber \\
     &+&\{\frac{8}{\pi}+\frac{4}{\pi}(n-1)(1-2\sum_{r=1}^{n/2}\frac{1}{2r-1})\}\frac{m^2}{Q^2}+O(\frac{m^4}{Q^4}\ln{\frac{Q^2}{m^2}}) \nonumber \\
&-&(Q^2\rightarrow\mu^2)
\end{eqnarray}
 The two-loop graphs depicted in Fig(3b,3c) have a leading asymptotic behavior $\simeq\frac{1}{Q^2}$,
corresponding to dimension 2 operators in an operator product expansion. The self-energy (Fig.3b)
and exchange (Fig.3c) graphs are not well-defined individually in light-cone gauge until a regularization
procedure has been given for the gluon propagator. We shall take the momentum space
gluon propagator to be $P(\frac{1}{k_{-}^2})\equiv {\rm Re}(\frac{1}{k_{-}+i\epsilon})^2$, where the
infinitesimal $\epsilon$ can only be sent to zero after combining graphs (3b) and (3c). 

\begin{figure}[htp]
\hbox to \hsize{\hss\psfig{figure=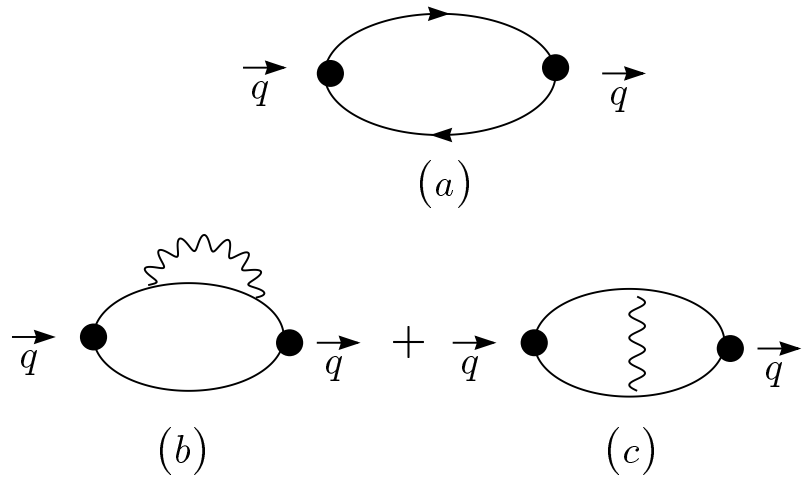,width=0.8\hsize}\hss}
\caption{One and two-loop graphs contributing to the correlators $M_n(Q^2)$}
\label{fig:pert}
\end{figure}

 A lengthy but straightforward evaluation of the two-loop Feynman integrals for the self-energy
graph Fig(3b) yields (for $n=0$) the asymptotic behavior
\begin{equation}
\label{eq:self}
 M_{0}^{2b}(Q^2)\simeq \frac{2}{\pi}(\frac{g^{2}N}{\pi})(\frac{1}{m^2}\ln(\epsilon)+\frac{1}{m^2}\ln(\frac{Q^2}{m^2})
-\frac{2}{Q^2})+O(\frac{1}{Q^4}\ln(\frac{Q^2}{m^2})^2)
\end{equation}
while the exchange graph yields
\begin{eqnarray}
\label{eq:exchange}
M_{0}^{2c}(Q^2)&\simeq&     
-\frac{2}{\pi}(\frac{g^{2}N}{\pi})(\frac{1}{m^2}\ln(\epsilon)+\frac{1}{m^2}\ln(\frac{Q^2}{m^2})
+\frac{1}{2m^2})  \nonumber \\
    &+&O(\frac{1}{Q^4}\ln(\frac{Q^2}{m^2})^2)
\end{eqnarray}
As expected, the regularization dependence cancels between the two graphs. The discontinuity
of this corrrelator is infrared-safe so the mass-singularities in $Q^2$ dependent terms must
also cancel, as they do, thereby removing the logarithmic terms completely. The power
singularity $\frac{1}{2m^2}$ in the exchange graph is not physical and is in fact removed
by the overall subtraction at $Q^2=\mu^2$ needed to define the overall amplitude. The total
2-loop contribution to this moment is thus (now, and henceforth, using units where $\frac{g^2 N}{\pi}=1$)
\begin{equation}
 M_{0}^{2-{\rm loop}}(Q^2)\simeq -\frac{4}{\pi}\frac{1}{Q^2}+O(\frac{1}{Q^4})
\end{equation}
The higher moments (we shall only need $n=2,4$) may be similarly computed and one finds in
all cases a pure power dependence
\begin{equation}
M_{n}^{2-{\rm loop}}(Q^2) \simeq \frac{2}{\pi}(n-2)\frac{1}{Q^2}+O(\frac{1}{Q^4})
\end{equation}
Since we are now in possession of the full $\frac{1}{Q^2}\ln(Q^2)$ contribution (arising solely from
the one loop graph Fig(3a)) the subdominant asymptotic coefficients $A_n$ (such that 
$f_{nk}\propto 1+\frac{A_{n}}{k}$ for large $k$) are now determined 
\begin{eqnarray}
  \frac{4}{\pi}(\gamma -1-\frac{\pi^2}{2}(A_{0}+A_{n}))&=&\frac{4}{\pi}(n-1)\gamma  \nonumber \\
  \Rightarrow \;\;A_{n}&=&\frac{1}{\pi^2}(2(1-n)\gamma -1)
\end{eqnarray}
Thus $A_{2}=A_{0}-\frac{4\gamma}{\pi^2}$, in agreement with the exact identity (\ref{eq:identity2}).

    Combining one and two-loop contributions, the asymptotic behavior of $M_n(Q^2)$, through
next to leading terms, is thus (after the overall subtraction at $Q^2=\mu^2$) given by
\begin{eqnarray}
 M_n(Q^2)&\simeq&\frac{2}{\pi}\ln(\frac{Q^2}{\mu^2})+\frac{4}{\pi}\frac{\gamma}{Q^2}\ln(\frac{Q^2}{\mu^2}) \nonumber \\ &+&\frac{2}{\pi}\{n-2+4\gamma+2(n-1)\gamma\ln(\frac{\mu^2}{\gamma})+2(n-1)\gamma(1-2\sum_{r=1}^{n/2}\frac{1}{2r-1})\}\frac{1}{Q^2}  \nonumber  \\ &+&O(\frac{1}{Q^4}\ln(\frac{Q^2}{\mu^2})^2)
\end{eqnarray}
where we remind the reader that the dimensionless variable $\gamma$ is simply the squared
quark mass $m^2$ in the natural units where $\frac{g^{2}N}{\pi}=1$. These analytic results serve as a
useful check on the numerical extraction of higher twist terms which we use below to implement the sum rules technique for the groundstate pion of this theory.

\section{Extraction of Moments by the Sum Rules Technique}

  In the standard sum rules approach to meson wavefunctions \cite{CZ}, the absorptive
part of a current-current correlator is modelled by a sum of resonances with
known mass values but unknown residues (typically at most the lowest two resonances
are used) at low $q^2$ values and by the perturbative QCD expression at large 
$q^2$. The Borel transform of this Ansatz (which is a smooth function, not
a distribution!) is then fitted to the corresponding
transform of an operator product expansion (OPE) for the same correlator
including higher twist terms over a suitable range of the Borel variable. 
In 4 dimensional QCD two sets of higher twist operators are included, 
so that terms falling off like $1/Q^4,1/Q^6$ are present explicitly (a potential $1/Q^2$ contribution is assumed absent, which is a source of some controversy, to which we
return below 
\cite{BYZ,DP}). The
hadronic matrix elements appearing in the OPE are estimated phenomenologically.
We shall follow the same procedure in 2-dimensional QCD. The only difference is
the inevitable presence of $1/Q^2$ terms arising from twist 2 operators, so that the inclusion of next to leading
and next to next to leading terms corresponds in this case to $1/Q^2$ and
$1/Q^4$ behavior.
\begin{figure}[htp]
\hbox to \hsize{\hss\psfig{figure=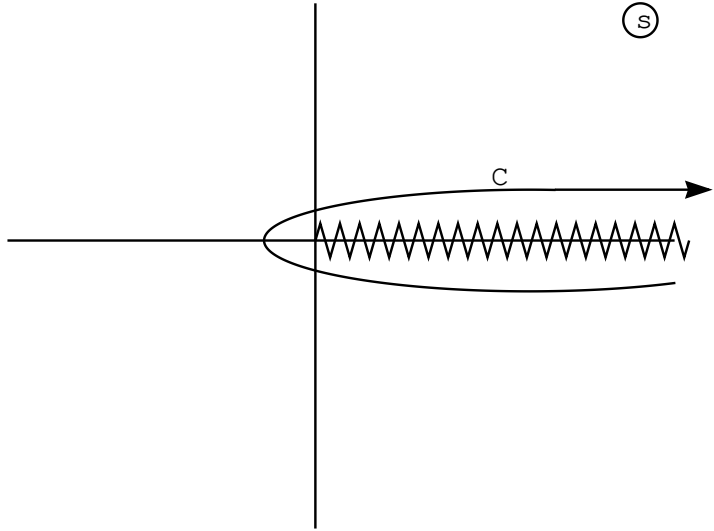,width=0.8\hsize}\hss}
\caption{Contour defining Borel Transform}
\label{fig:fig4)}
\end{figure}

  Define $s\equiv q^2$ ($>0$ in the timelike regime) and the Borel transform
of $f(s)$ as
\begin{equation}
\label{eq:BT}
 \tilde{f}(M^2)\equiv\frac{-1}{2\pi iM^2}\int_{C} f(s)e^{-s/M^2}ds
\end{equation}
with the contour $C$ as indicated in Fig(4). Thus the Borel transforms
of $\{\ln(-s),\frac{1}{s}\ln(-s),\frac{1}{s},\frac{1}{s^2},\rm etc\}$
are $\{1,\frac{\ln(M^2)-\gamma_e}{M^2},\frac{1}{M^2},-\frac{1}{M^4},\rm etc\}$.
Here $\gamma_e$ is the Euler constant. Writing (in the spacelike region
$-q^2=Q^2>0$)
\begin{eqnarray}
\label{eq:asexp}
 M_n(Q^2)&=&\frac{2}{\pi}\ln(\frac{Q^2}{\mu^2})-\{\frac{A_n}{Q^2}\ln(\frac{Q^2}{\mu^2})
 +\frac{B_n}{Q^2} \nonumber \\
  &+&\frac{C_n}{Q^4}\ln^2(\frac{Q^2}{\mu^2})+\frac{D_n}{Q^4}\ln(\frac{Q^2}{\mu^2}) 
+\frac{E_n}{Q^4} \nonumber \\
 &-&(Q^2\rightarrow\mu^2)\}+O(1/Q^6)
\end{eqnarray}
The corresponding Borel Transform is given by
\begin{eqnarray}
\label{eq:BTOPE}
 \tilde{M}_n(M^2)&=&\frac{2}{\pi}+\frac{A_n}{M^2}(\ln(\frac{M^2}{\mu^2})-\gamma_e)+\frac{
B_n}{M^2} \nonumber \\
 &+&\frac{C_n}{M^4}(1-\frac{\pi^2}{6}+(\ln(\frac{M^2}{\mu^2})+1-\gamma_e)^2)
+\frac{D_n}{M^4}(\ln(\frac{M^2}{\mu^2})+1-\gamma_e)+\frac{E_n}{M^4} \nonumber \\
&+&O(1/M^6)
\end{eqnarray}
On the other hand, splitting $M_n(s)$ into a contribution from $N$ low-lying
resonances (with squared masses below some cutoff $S_n$) and a high energy
perturbative piece $\theta(s-S_n)\cdot \frac{2}{\pi}\ln(\frac{-s}{\mu^2})$,
one may also write
\begin{equation}
\label{eq:BTDUAL}
 \tilde{M}_n(M^2)\simeq \frac{2}{\pi}e^{-S_n/M^2}+\frac{1}{M^2}
\sum_{k=1}^{k=N}\rho_{nk}e^{-\mu_{k}^{2}/M^2},\;\;\mu_N^2<S_n 
\end{equation}
\begin{equation}
\label{eq:residue}
  \rho_{nk}\equiv f_{nk}f_{0k} 
\end{equation} 
 The duality cut variable $S_n$ is allowed to float and is determined
in the fitting procedure when the right hand sides of (\ref{eq:BTOPE})
and (\ref{eq:BTDUAL})
 are matched, as are the residues $\rho_{nk}$.  We shall take
$N$=2 and fix the masses of the two lowest pseudoscalars at their
exact values obtained by diagonalizing the 't Hooft Hamiltonian (\ref{eq:Ham}).
Information about the shape of the groundstate meson (``pion") wavefunction
is obtained from the ratios $\frac{\rho_{21}}{\rho_{01}}=\frac{f_{21}}{f_{01}}$
 and $\frac{\rho_{41}}{\rho_{01}}=\frac{f_{41}}{f_{01}}$.

   We first examine the accuracy of the sum rules procedure  on the assumption
that the high momentum  behavior of  $M_n(Q^2)$ is known very accurately 
over a range of $Q^2$ where a sizable fraction of the variation is due to 
the higher twist contributions.  Specifically, the procedure involves the
following steps:\\
\begin{enumerate}
\item  The resonance sum representation is used to calculate $M_n(Q^2)$
 to high accuracy over a range  $\mu^2 <Q^2<200$, where the subtraction point 
has been chosen throughout as $\mu^2=5$ (in units where the squared mass
scale  intrinsic to the theory, namely $\frac{g^2N}{\pi}$ is set to unity).
For example, for the quark mass value $m_q^2=0.534$,  with $\mu_1^2=4.59$,
 we take $\mu^2=5$.  The resulting values are then fit over a range $Q_0^2<Q^2<200$
 (there is almost no dependence on the upper cutoff, once it is chosen 
reasonably large) 
to the asymptotic expansion (\ref{eq:asexp}), allowing us to extract the coefficients $A_n,B_n,...E_n$.
The low point  of the fit range $Q_0^2$ is adjusted to obtain the best  fit , and typically
one  finds $Q_0^2\simeq$20-30, with a rms deviation $<10^{-5}$.  As a check that
this procedure is yielding sensible results, we can compare the coefficients  $A_n,B_n$
 obtained from the fit with the analytic values obtained in the preceding section.
 The results for $m_q^2=$0.534 are shown in the top of Table 1,  for moments $n=$0,2,4, with
 the analytic (i.e. large N) values in parentheses.  Evidently, this fitting procedure reproduces
the twist 2 coefficients to about 10\%.  We shall argue below that the precise values of the
coefficients are in fact not  too important, as long as one  has a good uniform fit to the correlators
$M_n$ over a large enough $Q^2$ range.
\item  Once the coefficients $A_n,B_n,..,E_n$ are known the Borel transform (\ref{eq:BTOPE})
 can be computed over any given range $M_0^2<M^2<200$.  Then a fit is performed
 by matching the Ansatz 
 (\ref{eq:BTOPE}) to (\ref{eq:BTDUAL}) over a range of  $M^2$ values where the
 higher twist contribution is appreciable. A convenient choice yielding
perfectly reasonable results is 
 to take $M_0^2\simeq \mu^2 (=5)$.  The fitting parameters in the
  Ansatz (\ref{eq:BTDUAL}) are $S_n$, the duality cut, and the residues of the two lowest mesons
 $\rho_{n1},\rho_{n2}, n=0,2,4$.  From the residues $\rho_{nk}$  one  easily solves 
 for the moments $f_{nk}$ (cf (\ref{eq:residue})).  The results for the moments are  shown in the last column of  Table 1,
 together with the exact values in parentheses.  The first three moments are reproduced
 fairly well, to 10\% accuracy.  The same procedure applied to a heavier mass quark
(``onium" type meson, with $m_q^2=$2.6) leads to OPE coefficients and moments 
shown at the bottom of Table 1. The first two moments are still given reasonably well, but the $n=4$
moment is too large.
\end{enumerate}

\begin{table}
\begin{center}
\begin{tabular}{|c|c|c|c|c|c|c|} 
\multicolumn{7}{c}{Quark mass (squared) = 0.534} \\
\hline
\multicolumn{1}{|c|}{n (Moment)}
&\multicolumn{1}{c|}{$A_n$}
&\multicolumn{1}{c|}{$B_n$}
&\multicolumn{1}{c|}{$C_n$}
&\multicolumn{1}{c|}{$D_n$}
&\multicolumn{1}{c|}{$E_n$}
&\multicolumn{1}{c|}{$f_{n1}$}  \\ \hline
0  &  0.596(0.680)  & 2.02(2.114) & -0.398  & 2.51 & 1.98 & 3.47(3.41) \\
2  &  -0.775(-0.680) & -1.948(-2.201) & 3.303 & 4.742 & 1.269 & 1.60(1.82) \\
4  &  -2.11(-2.040) & -3.31(-3.796) & 3.81 & 6.92 & 4.56 & 1.37(1.40) \\
\hline
\multicolumn{7}{c}{Quark mass (squared) = 2.60} \\
\hline
0  &  3.27(3.31)    & 0.61(0.13)  & -1.84   & -13.7& -2.89& 4.87(4.83) \\
2  &  -3.69(-3.31)   & -4.17(-5.48)   & -3.06 & 1.93  & 4.11  & 1.75(1.58) \\
4  &  -9.58(-9.93)  & 2.95(2.16)   & -47.4& -18.9& 0.83 & 1.61(0.97) \\
\hline
\end{tabular}
\caption{Asymptotic coefficients and moments (exact values in parentheses)}
\end{center}
\end{table}
   At this point it is appropriate to point out that accurate control of 
logarithmic terms in the asymptotics
of $M_n(Q^2)$ is {\em not} essential for this procedure to give reasonably accurate results. Indeed, the
 Borel Transform (\ref{eq:asexp}) to (\ref{eq:BTOPE}) is evidently a linear mapping which becomes a bounded 
 one (relative
 to the $L^2$ norm, which is relevant here as we perform least square fits throughout)
 once finite intervals in $Q^2$ and $M^2$,
 and a finite truncation of the OPE, are chosen.  Thus the results obtained for
 $\rho_{nk}$ are really quite insensitive to the precise values of the coefficients
$A_n,..E_n$, provided only that the asymptotic form is a good representation in the
mean of the exact
 $M_n(Q^2)$ over  a suitable  $Q^2$ range.  For example, if we use a pure power fit for
the higher twist terms
\begin{equation}
  M_n(Q^2) \simeq \frac{2}{\pi}\ln(\frac{Q^2}{\mu^2})-\{\frac{A_n^{\prime}}{Q^2}
 +\frac{B_n^{\prime}}{Q^4}+\frac{C_n^{\prime}}{Q^6}+\frac{D_n^{\prime}}{Q^8}\}
\end{equation}
 which gives a fit with rms deviation $<10^{-3}$ in the range  $\mu^2<Q^2<200$
(considerably worse than the fit obtained through twist 4, but including logarithmic
 terms), one finds $f_{01}=$3.38, $f_{21}=$1.95, and $f_{41}=$1.70.  Only the last
 moment , n=4, seems to suffer (to the extent of  about a 20\% error) from the use
 of a pure power Ansatz for higher twist terms. Typically the QCD applications\cite{CZ}   employ only pure 
power dependencies in higher twist terms. 

\section{Truncation Sensitivity of the Method}

   We saw previously (cf Section 2) that as simple a modification of the structure
of the theory as the introduction of a lattice high-momentum cutoff can lead to 
a profound modification of the form of the ground-state meson wavefunctions. 
In the language of the Borel transform, the effect of such a cutoff is readily determined. 
A short calculation shows that the one-loop contribution to the Borel transform 
$\tilde{M}_{n}(M^2)$ in the presence of a UV cutoff $\Lambda=\pi/a$ ($a$ the
lattice spacing) is given by
\begin{equation}
\label{eq:discbor}
\tilde{M}_{n}^{\rm 1-loop}(M^2)=\frac{8}{\pi M^2}\int_{-\pi/a}^{\pi/a}
e^{-4E_{k}^2/M^2}E_{k}^{1-n}(E_{k}^2-m^2)^{n/2}dk
\end{equation}
where
\begin{equation}
\label{eq:latten}
  E_{k}\equiv\sqrt{\frac{4}{a^2}\sin^{2}(\frac{ak}{2})+m^2}
\end{equation}
In the continuum limit $a\rightarrow 0$  the integral (\ref{eq:discbor}) becomes
divergent for $M^2\rightarrow\infty$ and the large $M^2$ behavior is amplified
to a constant term. For example, $\tilde{M}_{0}(M^2)$ is given explicitly
in the continuum limit by $\frac{2}{\pi}\frac{2m^2}{M^2}e^{-2m^2/M^2}(K_0(\frac{2m^2}{M^2})
+K_1(\frac{2m^2}{M^2}))\rightarrow\frac{2}{\pi},\;\;M^2\rightarrow\infty$.  By
contrast, in the presence of the UV lattice cutoff, the leading ``twist-0" term is
removed entirely, and the asymptotic behavior begins at order $1/M^2$. So we can
certainly expect trouble whenever the balance of lower and higher twist terms 
is altered.

  We saw in the preceding section that  the sum rules method appears to give reasonable results once a uniformly
accurate fit to the $M_n$ correlators is given. We shall now show directly that
 it  can fail quite substantially if the
balance between the various higher twist nonperturbative contributions is altered.
In the applications of the method to 4-dimensional QCD \cite{CZ}, the overall size of these
higher twist  terms is determined by a  phenomenologically estimated condensate
(i.e. expectation of a hadronic composite operator), while the moment dependence 
is obtained from a perturbatively computed coefficient function. Assuming the
absence of the first infrared renormalon (on which more below) the twist 4 and 6
contributions are estimated from sum rules for charmonium \cite{charm} and from PCAC \cite{PCAC}.
Thus different systematic errors are possible in different higher twist terms. Moreover,
as these condensates involve the scale of the theory $\Lambda_{QCD}$ to high powers,
a relatively small ambiguity in the scale of the theory can alter the balance of 
higher twist terms quite substantially.  We can examine the effect of a similar systematic
error in the 2-dimensional model by applying a scale factor to the twist 4 term, 
leaving the perturbative and twist 2 terms unchanged.   Imitating the procedure
used for QCD as closely as possible we first extract the best fit to the exact correlators
$M_n(Q^2)$  (for $m_q^2=$0.534) using pure powers only (through $1/Q^4$), and then plot the sensitivity
of the ratios $f_{21}/f_{01},f_{41}/f_{01}$ to an overall rescaling of the twist 4
contribution (see Fig. 5). Evidently, the higher moments increase steadily with the overall
scale of the twist 4 term. In the QCD case,  the considerably higher values found
for the second and fourth moments (as compared to the asymptotic wavefunction
$\propto x(1-x)$) were interpreted by Chernyak and Zhitnitsky 
\cite{CZ} as evidence for a
non-convex, doubly peaked wavefunction. 

\begin{figure}[htp]
\hbox to \hsize{\hss\psfig{figure=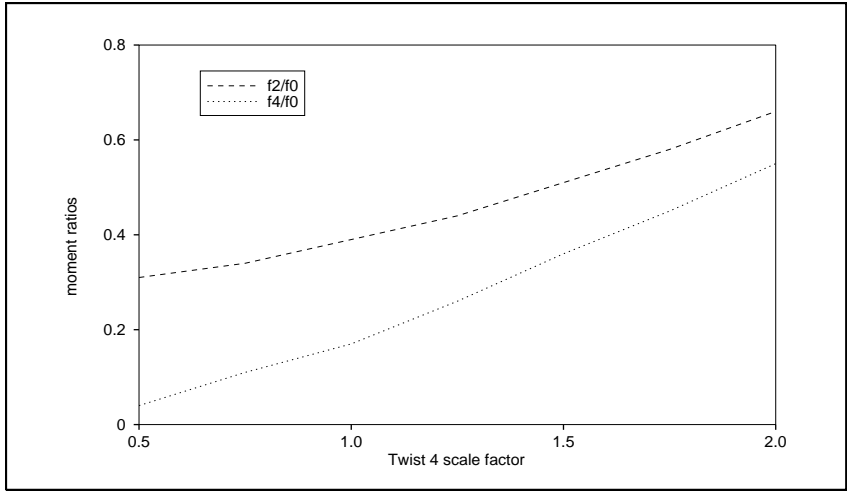,width=0.8\hsize}\hss}
\caption{Sensitivity of moments to uniform rescaling of twist-4 terms}
\label{fig:fig5)}
\end{figure}

  It was recently argued  \cite{BYZ,DP} that there is no known rigorous argument to exclude the 
presence of an intrinsically nonperturbative $1/Q^2$ contribution to the
coefficient function of the identity operator,  and hence to the correlators
$M_n(Q^2)$ (this issue is intimately related with the location of the first infrared
renormalon in the theory).  In fact, if we return to the full  logarithmic fits summarized
in Table 1, and assume ignorance of the twist 2 terms by setting the coefficients
$A_n,B_n$=0,  we find {\em perfectly consistent} fits to the duality Ansatz (\ref{eq:BTDUAL}). It should
be emphasized that the fitting procedure does not provide any {\em internal}
evidence  that important contributions are missing. Indeed, the quality of the fits
(using rms deviation as a figure of merit) actually improves by almost an order of magnitude!
And the optimal fits find the duality cut variable (which is allowed to float in the
fitting procedure) $S_n$ settling  at a self-consistent value, namely $\mu_2^2<S_n<\mu_3^2$
(since the lowest two mesons are included explicitly in the resonance sum part).  The effect on the moments however is dramatic, and goes in the same direction as an increase in twist  4 contributions relative
to lower twist terms discussed above.  One now finds $f_{21}/f_{01}\simeq 1.0$ and
$f_{41}/f_{01}\simeq 1.1$ implying a wavefunction with a minimum at $x=1/2$, rather than
the correct convex and singly peaked result. 

  As regards the implications of our results for 4 dimensional QCD, it must be admitted
that the prospects for a really firm nonperturbative resolution of the pion
wavefunction seem rather dim at present. From the perspective of QCD sum rules,
a reliable result would seem to depend on reasonably accurate uniform control
over the correlators over a wide $Q^2$ range. First principles calculations 
starting from lattice QCD, on the other hand, also seem to introduce truncations
of the theory which lead to potentially dangerous distortions of wave-function
structure. Further calculations to study the systematic lattice errors in both
correlators and moments on 4 dimensional (quenched) lattices of the size presently available
are in progress.

\vspace{1in}
 \section{Acknowledgements}
 A.D. acknowledges gratefully the support of the EKA Fund, Columbia University,
and of the NSF through grant 93-22114. S.P. was supported in part through 
Department of Energy Grant No. DE-FG02-91ER40685. E.S. was supported in
part through DFG contract Li519/2-1.

\end{document}